\documentstyle[preprint,prl,aps]{revtex}

\begin{document}
\preprint{\vbox{\hbox{February 1995}\hbox{rev. May 1995}\hbox{IFP-713-UNC}}}
\draft
\title{Quark Mass Textures within a Finite Non-abelian Dicyclic Group.}
\author{\bf Paul H. Frampton and Otto C. W. Kong}
\address{Institute of Field Physics, Department of Physics and Astronomy,\\
University of North Carolina, Chapel Hill, NC  27599-3255}
%\date{\today}
\maketitle

\begin{abstract} Using as a flavor symmetry a finite nonabelian dicyclic
$Q_{2N}$ group, we show how to derive quark
mass matrices with two arrangements of symmetric texture zeros which are
phenomenologically viable. Three other such
acceptable textures in the recent literature are unattainable in this
approach and hence disfavored. We assume massive vector-like
fermions and Higgs singlets transforming as
judiciously-chosen $Q_{2N}$ doublets and use the tree-level mass
generation mechanism of Froggatt and Nielsen.

\end{abstract}

\newpage
One of the longest standing problems in particle theory has been the values
of the fermion masses. Aside from the compelling classification of
quarks and leptons into families, based on  masses and  quark
mixing angles, the masses themselves remain a  cunundrum.
For example, the mass hierarchy: why is the top quark over a third of a million
times heavier than the electron?

There has
recently been considerable interest in the structure of the quark mass
matrices, particularly in the idea of postulating texture zeros in grand
unified theories\cite{GUTS1},\cite{GUTS2},\cite{GUTS3} with a view to
obtaining relations
between the masses and mixing angles.  A list of  phenomenologically
viable
quark mass matrices bearing  a maximum number of symmetric
texture zeros
 was presented in\cite{RRR}. The possibility of constructing such
mass matrices from a
scheme of gauged flavor symmetry has been
considered in \cite{U1P} and \cite{U1}.

There has also been considerable activity in the use of finite non-abelian
groups as flavor
symmetries\cite{finite1},\cite{finite2},\cite{finite3}, with a view to
generating the mass hierarchy.

Here we attempt a
synthesis of these two approaches and construct desirable quark mass
matrix
textures by using a nonabelian flavor symmetry (specifically a dicyclic
group $Q_{2N}$) together with the Froggatt-Nielsen mechanism of mass
generation\cite{FN}. While U(1) flavor symmetry constructions for quark mass
matrices
with {\it nonsymmetric} hierarchical textures have  been attempted
\cite{nonsym},  the full list of such phenomenologically viable quark mass
matrices
is not yet available. Our approach does not {\it a priori} give a symmetric
texture. However, as a first attempt, we consider here only the possibility
of constructing the symmetric texture patterns presented in [4]. The general
case of non-symmetric textures would  naturally be a very interesting next
step.

The use of $Q_{2N}$ as a finite flavor group has been discussed in detail in
\cite{finite3}. We recall here that the irreducible representations of $Q_{2N}$
are four
singlets $1, 1^{'}, 1^{''}, 1^{'''}$ and $(N - 1)$ doublets $2_{(k)}$,
with $1 \leq k \leq (N - 1)$. Most important for our purposes are the
products:
\begin{equation}
2_{(k)} \times 2_{(l)} = 2_{(|k-l|)} + 2_{(min\{k+l,2N-k-l\})}
\end{equation}
where, in a generalized notation, $2_{(0)} \equiv 1 + 1^{'}$ and $2_{(N)}
\equiv 1^{''} + 1^{'''}$.

We assign the quarks to $Q_{2N}$ representations as follows:
$$\begin{array}{cccc}
\left. \begin{array}{c} \left( \begin{array}{c} t \\ b \end{array} \right)_{L}
\\
\left( \begin{array}{c} c \\ s \end{array} \right)_{L} \end{array}  \right\} &
2_{(2)}&
 \begin{array}{c} \left. \begin{array}{c} t_R\\c_R \end{array} \right\} \\
\left. \begin{array}{c} b_R\\s_R \end{array} \right\} \end{array}&
\begin{array}{c} 2_{(2)}\\
2_{(1)} \end{array} \\
\left( \begin{array}{c} u \\ d \end{array} \right)_{L}  &
1^{'}&\begin{array}{c} u_R\\d_R
\end{array} & \begin{array}{c} 1^{'}\\1 \end{array} \\
\end{array}$$

When we embed the finite spinorial $Q_{2N}$ into its continuous progenitor
$SU(2)$,
$1, 2_{(1)}$ and $(1^{'} +2_{(2)})$ correspond respectively to the singlet,
doublet and triplet representations. The above quark assignment is thus
anomaly-free if the leptons are assigned to: $$\begin{array}{cccc}
\left. \begin{array}{c} \left( \begin{array}{c} \nu_{\tau} \\ \tau^{-}
\end{array} \right)_{L} \\
\left( \begin{array}{c} \nu_{\mu} \\ \mu^{-} \end{array} \right)_{L}
\end{array}  \right\} & 2_{(1)}&
\left.  \begin{array}{c}  \tau^{+}_L\\ \mu^{+}_L \end{array} \right\}
&  2_{(2)} \\
\left( \begin{array}{c} \nu_{e} \\ e^{-} \end{array} \right)_{L}  & 1& e^{+}_L
 & 1^{'} \\
\end{array}$$

We shall not consider lepton masses further here. For the mass textures of the
quarks we postulate heavy
vector-like fermions and singlet Higgs and assume the quark masses
arise from tree graphs as in \cite{FN}.

As the first of two successful examples, we demonstrate how to
derive the five-zero texture in Eqs. (2) and (3) below. [Note that no texture
with the
maximum number of six texture zeros can be phenomenolically viable\cite{RRR}.]

\begin{equation}
$$M_{u} = \left(\begin{array}{ccc}
0 & 0 & \lambda^{4} \\
0 & \lambda^{4} & \lambda^{2} \\
\lambda^{4} & \lambda^{2} & 1
\end{array}\right)$$
\end{equation}
\begin{equation}
$$M_{d} = \left(\begin{array}{ccc}
0 & \lambda^{4} & 0 \\
\lambda^{4} & \lambda^{3} & 0 \\
0 & 0 & 1
\end{array}\right)$$
\end{equation}

For these matrices we have suppressed all coefficients of order one since at
the present
stage we are satisfied to derive only the correct orders in
$\lambda$ for each entry.

The standard Higgs scalar doublets of $SU(2)_L$ are taken as a $2_{(4)}$ of
$Q_{2N}$ coupling
to the up quarks, and a $2_{(3)}$ coupling to the down quarks. We assume
these get VEVS that break $Q_{2N}$ and give mass only to the third family. For
the up quark mass matrix the entry $(M_u)_{33}$ is
of order 1 from the coupling $t_L(2_{(2)})t_R(2_{(2)})H_u(2_{(4)})$. At
leading order, all other entries $(M_u)_{ij}$ vanish. Similarly,
$b_L(2_{(2)})b_R(2_{(1)})H_d(2_{(3)})$ gives $(M_d)_{33}$ of order 1 and no
other $(M_d)_{ij}$.

To obtain the other entries in Eqs. (2) and (3) at order $\lambda ^n
(\lambda \sim sin\theta_{C} \sim 0.22$
where $\theta_{C}$ is the Cabibbo angle), we
introduce a list of vector-like quark doublets
$ Q_i(2_{(i)}) (i=6, 7, 10, 13, 14),$
singlets
$U_i(2_{(i)}) (i=6, 10, 14)$
and
$D_i(2_{(i)}) (i=4, 17),$
 bearing the same standard model quantum numbers as
 $Q_L, u_R$ and $d_R$ respectively, together with standard model
singlet Higgses
$S_i(2_{(i)}),(i= 5, 8, 13, 14, 17, 20)$. Although this set of $Q_{2N}$
doublets seems long and {\it ad hoc}, it is highly constrained (see below).
Since we have assumed heavy particles in doublets up to $2_{20}$ the
flavor group, of order 84, is $Q_{42}$.

We choose a set of bases and label the two states in the heavy fermion $Q_{2N}$
doublets as $2_{(i)+}$ and $2_{(i)-}$, which lie
respectively in the third and second family direction.  The
$H_u$ VEV then allows only the six couplings:
\[	t_L<H_u>U_{6+}; \; Q_{6+}<H_u>t_R; \; Q_{6+}<H_u>U_{10+}; \]
\[  Q_{10+}<H_u>U_{6+};\; Q_{10+}<H_u>U_{14+}; \; Q_{14+}<H_u>U_{10+}.  \]
%\[ t_L<H_u>U_{6+}, \; Q_{6+}<H_u>t_R, \; Q_{6+}<H_u>U_{10+},
%\; Q_{10+}<H_u>U_{6+};\]
and the $H_d$ VEV only the two couplings:
\[	Q_{7+}<H_d>D_{4+};\; Q_{14+}<H_d>D_{17+}.\]
The $S_i$ VEVS may then be chosen to give certain vertices such as :
$U_{6+}^{\dag}<S_{8+-}>c_R,$ \\   $U_{10+}^{\dag}<S_{8+-}>c_R$, and others.
% \; U_{6+}^{\dag}<S_{8+-}>U_{14-}, \; U_{14-}^{\dag}<S_{14-}>u_R;
%\]
%\[ 	U_{6+}^{\dag}<S_{8+-}>c_R,  \; U_{10+}^{\dag}<S_{8+-}>c_R,
% \; U_{10+}^{\dag}<S_{8+-}>U_{18-}, \; U_{18-}^{\dag}<S_{18-}>u_R;\]
%\[ 	D_{4+}^{\dag}<S_{5+-}>s_R, \; D_{4+}^{\dag}<S_{13+-}>D_{17-},
%\; D_{17-}^{\dag}<S_{17-}>d_R;
%\]
%\[ 	D_{4+}^{\dag}<S_{5+-}>s_R, \; D_{4+}^{\dag}<S_{15+-}>D_{11-},
%\; D_{11-}^{\dag}<S_{11-}>d_R; \]
%\[ 	Q_{6+}^{\dag}<S_{8+-}>c_L,  \; Q_{10+}^{\dag}<S_{8+-}>c_L,
% \; Q_{6+}^{\dag}<S_{8+-}>Q_{14-}, \; Q_{14-}^{\dag}<S_{14-}>u_L;
%\]
%\[
%Q_{7+}^{\dag}<S_{5+-}>s_L, \; Q_{7+}^{\dag}<S_{20+-}>Q_{13-}, \;
%Q_{13-}^{\dag}<S_{13-}>d_L.
%\]
%\[ Q_{6+}^{\dag}<S_{8+-}>c_L, \; Q_{10+}^{\dag}<S_{8+-}>c_L,
% \; Q_{10+}^{\dag}<S_{8+-}>Q_{18-}, \; Q_{18-}^{\dag}<S_{18-}>u_L;\]
%\[ Q_{7+}^{\dag}<S_{5+-}>s_L, \; Q_{7+}^{\dag}<S_{16+-}>Q_{23-}\]
%we use $c_L$ and $s_L$ to denote the same $SU(2)_L$ doublet, and likewise
%for $u_L$ and $d_L$.
We define:
\begin{equation}
\lambda^2 = \frac{<S_i>}{M_{even}}
\end{equation}
\begin{equation}
\lambda = \frac{<S_i>}{M_{odd}}
\end{equation}
where $M_{even}$ and $M_{odd}$ denote
the mass of a heavy fermion in $Q_{2N}$ representation $2_{(k)}$ for being
$k$ even and odd respectively. Note that Eqs. (4) and (5) are acceptable
because
the k-even and
k-odd doublets occur independently in the irreducible representations of
the covering $SU(2)$ in the sense that the k-even doublets appear only in
vectors of SU(2) and the k-odd doublets appear only in spinors.

We have now all the ingredients of the model. In the low energy efffective
field theory, after integrating out the heavy fermions\cite{EFT}, we have
the tree level quark mass matrices having the structure of the
model denoted by roman numeral V in ref.\ \cite{RRR}, namely those exhibited in
Eqs.\~(2)
and (3) above.

For instance, the  $(M_u)_{32}$ entry is given by the Froggatt-Nielsen
tree graph (shown in Fig.(1a)) correponding to the operator couplings
\[  	t_L <H_u> ( U_{6+} U_{6+}^{\dag} ) <S_{8+-}> c_R
 = \lambda ^2 <H_u> t_L c_R;
\]
while $(M_u)_{13}$ is given by the graph of Fig. (1c) corresponding to:
\[ 	t_L <H_u> ( U_{6+} U_{6+}^{\dag} )
<S_{8+-}> (U_{14-}U_{14-}^{\dag})  <S_{14-}>u_R\]
\[ = \lambda ^4 <H_u>  t_L u_R; \]
and $(M_d)_{22}$ is given by Fig.\ (2a) corresponding to:
\[ 	s_L <S_{5+-}> (Q_{7+}^{\dag}Q_{7+}) <H_d> (D_{4+}
D_{4+}^{\dag})<S_{5+-}>s_R\]
\[ =  \lambda^3 <H_d> s_L s_R.  \]
The other entries in $M_u$ and $M_d$ are derived similarly; some further
examples
are shown in Fig. (1b) for $M_u$ and Figs. (2b) and (2c) for $M_d$.

In the construction  of the model, we followed a systematic procedure and were
surprised
to realize that it is highly non-trivial if any consistent model
can be constructed at all. The difficulty is not only to derive the correct
texture zeros but also to avoid unwanted entries at too low an order in
$\lambda$.
We find only two consistent models, the above model and one alternative
summarized below.

The mass matrices for the alternative model  have texture structures of the
model denoted by roman numeral IV in ref.\ \cite{RRR}. They are:

\begin{equation}
$$M_{u} = \left(\begin{array}{ccc}
0 & \lambda^{6} & 0 \\
\lambda^{6} & \lambda^{4} & \lambda^{2} \\
0 & \lambda^{2} & 1
\end{array}\right)$$
\end{equation}

\begin{equation}
$$M_{d} = \left(\begin{array}{ccc}
0 & \lambda^{4} & 0 \\
\lambda^{4} & \lambda^{3} & 0 \\
0 & 0 & 1
\end{array}\right)$$
\end{equation}

Note that the $M_{d}$ matrix is the same as in our first example but
$M_{u}$ is changed; as before, we neglect coefficients of order unity. The
$Q_{2N}$
assignments for the quarks and leptons are the same as they were previously.

The entries in Eqs.\ (6) and (7) at order $\lambda ^n$ are constructed, as in
the previous example, through introducing
 vector-like $Q_{2N}$ quark doublets
$Q_i(2_{(i)}) (i=6, 7, 10, 18, 23),$
singlets
$U_i(2_{(i)}) (i=6, 10, 18)$
and $D_i(2_{(i)}) (i=4,11),$
 together with standard model
singlet Higgses
$S_i(2_{(i)}),(i= 5, 8, 11, 15, 16, 18, 23)$. As mentioned in our first
example this set of doublets which seems long and {\it ad hoc} is really
highly constrained. For this second example, the flavor group is $Q_{48}$.

%We again choose a set of bases and label the two states in the heavy fermion
%%$Q_{2N}$
%doublets as $2_{(i)+}$ and $2_{(i)-}$, which lie
%respectively in the third and second family direction.  The
Under the same kind of state labelling, the $H_u$ VEV then allows only the four
couplings:
\[ t_L<H_u>U_{6+}, \; Q_{6+}<H_u>t_R,\]
\[ Q_{6+}<H_u>U_{10+}, \; Q_{10+}<H_u>U_{6+};\]
and the $H_d$ VEV only the coupling:
\[	Q_{7+}<H_d>D_{4+}.\]
The $S_i$ VEVS may then be chosen to give certain vertices such as :
$U_{6+}^{\dag}<S_{8+-}>c_R,\\
\; U_{10+}^{\dag}<S_{8+-}>c_R$, and so on.
% \; U_{6+}^{\dag}<S_{8+-}>U_{14-}, \; U_{14-}^{\dag}<S_{14-}>u_R;
%\]
%\[ 	U_{6+}^{\dag}<S_{8+-}>c_R,  \; U_{10+}^{\dag}<S_{8+-}>c_R,
% \; U_{10+}^{\dag}<S_{8+-}>U_{18-}, \; U_{18-}^{\dag}<S_{18-}>u_R;\]
%\[ 	D_{4+}^{\dag}<S_{5+-}>s_R, \; D_{4+}^{\dag}<S_{13+-}>D_{17-},
%\; D_{17-}^{\dag}<S_{17-}>d_R;
%\]
%\[ 	D_{4+}^{\dag}<S_{5+-}>s_R, \; D_{4+}^{\dag}<S_{15+-}>D_{11-},
%\; D_{11-}^{\dag}<S_{11-}>d_R; \]
%\[ 	Q_{6+}^{\dag}<S_{8+-}>c_L,  \; Q_{10+}^{\dag}<S_{8+-}>c_L,
% \; Q_{6+}^{\dag}<S_{8+-}>Q_{14-}, \; Q_{14-}^{\dag}<S_{14-}>u_L;
%\]
%\[
%Q_{7+}^{\dag}<S_{5+-}>s_L, \; Q_{7+}^{\dag}<S_{20+-}>Q_{13-}, \;
%Q_{13-}^{\dag}<S_{13-}>d_L.
%\]
%\[ Q_{6+}^{\dag}<S_{8+-}>c_L, \; Q_{10+}^{\dag}<S_{8+-}>c_L,
% \; Q_{10+}^{\dag}<S_{8+-}>Q_{18-}, \; Q_{18-}^{\dag}<S_{18-}>u_L;\]
%\[ Q_{7+}^{\dag}<S_{5+-}>s_L, \; Q_{7+}^{\dag}<S_{16+-}>Q_{23-}\]

The general procedure is as follows: the quarks and leptons has to come from
$SU(2)_H$ singlets, doublets and triplets. There are only 21 anomaly free
schemes of
assignment with no extra chiral fermions\cite{SU2H}. We are aiming at
assignments that
can lead to up- and down-quark mass matrices with different hierarchical
textures\cite{RRR}. That leaves us with two schemes of which only the one used
here gives
interesting models.

 Picking the above
scheme, the feasiblity of using the Froggatt-Nielsen
mechanism enforces the first family quarks to be $Q_{2N}$
singlets. We then introduce appropriate heavy fermions and Higges whenever
necessary as we go on to build entries of higher order in $\lambda$,
keeping track of overall consistency.

Attempts to construct models giving texture models I, II, and III of ref.
\cite{RRR} lead to conflicts, and we therefore conclude
that those patterns of texture zeros are disfavored.

In this approach, the standard model gauge group $G =SU(3)_C \times SU(2)_L
\times
U(1)_Y$ is extended to $G \times (Q_{2N})_{global}$ which, in turn,
is assumed a subgroup of $G \times SU(2)_{H}$ where $SU(2)_H$ is
a gauged horizontal symmetry. This last point is important because
the imposition of the necessary anomaly cancellation restricts the
assignment of the quarks and leptons to $Q_{2N}$ representations as discussed
above.

The authors of \cite{RRR} have analyzed all possible symmetric quark mass
matrices
with the maximal (six) and next-to-maximal (five) number of texture zeros, and
concluded that only five models, denoted by the roman numerals I to V in their
work,
are phenomenologically viable. By insisting on derivation of the texture zeros
from the $Q_{2N}$ dicyclic flavor symmetry, we have reduced the number of
candidates to
two, denoted in \cite{RRR} by the roman numerals IV and V.

It seems likely that similar considerations could be made directly at the
$SU(5)
\times Q_{2N}$
level because the assignment of the quarks and leptons are consistent with
an $SU(5)$ embedding, and this - together with the generalization to
supersymmetry -
are presently under investigation.

In conclusion, the reduction in the number of free parameters in the low energy
theory attained by postulating texture zeros in the fermion mass matrices
has been shown to have a dual description in terms of a horizontal
symmetry $Q_{2N} \subset SU(2)_H$. This $SU(2)_H$ could arise in a GUT group
or directly from a superstring. Our main point is that the derivation of
the values of the fermion masses and quark mixings in a putative
theory of everything may likely involve a horizontal
symmetry, probably gauged, as an important intermediate step. The two simple
cases given in the present paper
illustrate how this can happen.\\
\bigskip
\bigskip
\bigskip
\bigskip
\bigskip

This work was supported in part by the U.S. Department of
Energy under Grant DE-FG05-85ER-40219, Task B.\\

We thank an anonymous referee for helping to clarify the presentation. \\

\bigskip
\bigskip

{\bf Figure Captions.}\\

Fig.1  Froggatt-Nielsen tree graphs for $M_u$. (The symmetric counterpart
$(M_u)_{23}$,
and second graphs for $(M_u)_{22}$ and $(M_u)_{33}$ are not shown).\\

\bigskip

Fig.2  Froggatt-Nielsen tree graphs for $M_d$.

\newpage

\begin{flushleft}
\begin{figure}[h]

\vspace*{2.0cm}

\setlength{\unitlength}{1.0cm}

\begin{picture}(15,15)

\thicklines

\put(1,13.5){\framebox(2.3,1){$(a)$: $(M_u)_{32}$}}
\put(7.08,12){\vector(1,0){1}}
\put(8.08,12){\line(1,0){0.84}}
\multiput(9,12)(0,0.3){9}{\line(0,1){0.25}}
\put(8.85,14.6){${\bf \times}$}
\put(9.08,12){\line(1,0){0.84}}
\put(10.92,12){\vector(-1,0){1}}
\multiput(11,12)(0,0.3){9}{\line(0,1){0.25}}
\put(10.85,14.6){${\bf \times}$}
\put(11.08,12){\line(1,0){0.84}}
\put(12.92,12){\vector(-1,0){1}}
\put(6.5,11.9){$t_L$}
\put(9.2,14.6){$ \left \langle {H_u} \right \rangle $}
\put(10.2,11.5){$U_{6+}$}
\put(11.2,14.6){$ \left \langle {2_8} \right \rangle $}
\put(13,11.9){$c_R$}

\put(1,8.5){\framebox(2.3,1){$(b)$: $(M_u)_{22}$}}
\put(7.08,7){\vector(1,0){1}}
\put(8.08,7){\line(1,0){0.84}}
\multiput(9,7)(0,0.3){9}{\line(0,1){0.25}}
\put(8.85,9.6){${\bf \times}$}
\put(9.08,7){\line(1,0){0.84}}
\put(10.92,7){\vector(-1,0){1}}
\multiput(11,7)(0,0.3){9}{\line(0,1){0.25}}
\put(10.85,9.6){${\bf \times}$}
\put(11.08,7){\line(1,0){0.84}}
\put(12.92,7){\vector(-1,0){1}}
\put(4.5,6.9){$c_L$}
\put(9.2,9.6){$ \left \langle {H_u} \right \rangle $}
\put(10.2,6.5){$U_{6+}$}
\put(11.2,9.6){$ \left \langle {2_8} \right \rangle $}
\put(13,6.9){$c_R$}
\put(5.08,7){\vector(1,0){1}}
\put(6.08,7){\line(1,0){0.84}}
\multiput(7,7)(0,0.3){9}{\line(0,1){0.25}}
\put(6.85,9.6){${\bf \times}$}
\put(7.2,9.6){$ \left \langle {2_8} \right \rangle $}
\put(7.5,6.5){$Q_{10+}$}

\put(1,3.5){\framebox(2.3,1){$(c)$: $(M_u)_{13}$}}
\put(7.08,2){\vector(1,0){1}}
\put(8.08,2){\line(1,0){0.84}}
\multiput(9,2)(0,0.3){9}{\line(0,1){0.25}}
\put(8.85,4.6){${\bf \times}$}
\put(9.08,2){\line(1,0){0.84}}
\put(10.92,2){\vector(-1,0){1}}
\multiput(11,2)(0,0.3){9}{\line(0,1){0.25}}
\put(10.85,4.6){${\bf \times}$}
\put(11.08,2){\line(1,0){0.84}}
\put(12.92,2){\vector(-1,0){1}}
\multiput(13,2)(0,0.3){9}{\line(0,1){0.25}}
\put(12.85,4.6){${\bf \times}$}
\put(13.08,2){\line(1,0){0.84}}
\put(14.92,2){\vector(-1,0){1}}
\put(6.5,1.9){$t_L$}
\put(9.2,4.6){$ \left \langle {H_u} \right \rangle $}
\put(10.2,1.5){$U_{6+}$}
\put(11.2,4.6){$ \left \langle {2_8} \right \rangle $}
\put(15,1.9){$u_R$}
\put(12.2,1.5){$U_{14-}$}
\put(13.2,4.6){$ \left \langle {2_{14}} \right \rangle $}

\end{picture}

\caption{Froggatt-Nielsen tree graphs for $M_u$. (The symmetric
counterpart $(M_u)_{23},$ and second graphs for $(M_u)_{22}$ and
$(M_u)_{31}$ are not shown)}

\label{Fig. 1}

\end{figure}

\end{flushleft}

\newpage

\begin{flushleft}
\begin{figure}[h]

\vspace*{2.0cm}

\setlength{\unitlength}{1.0cm}

\begin{picture}(15,15)

\thicklines

\put(1,13.5){\framebox(2.3,1){$(a)$: $(M_d)_{22}$}}
\put(7.08,12){\vector(1,0){1}}
\put(8.08,12){\line(1,0){0.84}}
\multiput(9,12)(0,0.3){9}{\line(0,1){0.25}}
\put(8.85,14.6){${\bf \times}$}
\put(9.08,12){\line(1,0){0.84}}
\put(10.92,12){\vector(-1,0){1}}
\multiput(11,12)(0,0.3){9}{\line(0,1){0.25}}
\put(10.85,14.6){${\bf \times}$}
\put(11.08,12){\line(1,0){0.84}}
\put(12.92,12){\vector(-1,0){1}}
\put(4.5,11.9){$s_L$}
\put(9.2,14.6){$ \left \langle {H_d} \right \rangle $}
\put(10.2,11.5){$D_{4+}$}
\put(11.2,14.6){$ \left \langle {2_5} \right \rangle $}
\put(13,11.9){$s_R$}
\put(5.08,12){\vector(1,0){1}}
\put(6.08,12){\line(1,0){0.84}}
\multiput(7,12)(0,0.3){9}{\line(0,1){0.25}}
\put(6.85,14.6){${\bf \times}$}
\put(7.2,14.6){$ \left \langle {2_5} \right \rangle $}
\put(7.5,11.5){$Q_{7+}$}

\put(1,8.5){\framebox(2.3,1){$(b)$: $(M_d)_{21}$}}
\put(7.08,7){\vector(1,0){1}}
\put(8.08,7){\line(1,0){0.84}}
\multiput(9,7)(0,0.3){9}{\line(0,1){0.25}}
\put(8.85,9.6){${\bf \times}$}
\put(9.08,7){\line(1,0){0.84}}
\put(10.92,7){\vector(-1,0){1}}
\multiput(11,7)(0,0.3){9}{\line(0,1){0.25}}
\put(10.85,9.6){${\bf \times}$}
\put(11.08,7){\line(1,0){0.84}}
\put(12.92,7){\vector(-1,0){1}}
\put(4.5,6.9){$s_L$}
\put(9.2,9.6){$ \left \langle {H_d} \right \rangle $}
\put(10.2,6.5){$D_{4+}$}
\put(11.2,9.6){$ \left \langle {2_{13}} \right \rangle $}

\put(5.08,7){\vector(1,0){1}}
\put(6.08,7){\line(1,0){0.84}}
\multiput(7,7)(0,0.3){9}{\line(0,1){0.25}}
\put(6.85,9.6){${\bf \times}$}
\put(7.2,9.6){$ \left \langle {2_5} \right \rangle $}
\put(7.5,6.5){$Q_{7+}$}
\multiput(13,7)(0,0.3){9}{\line(0,1){0.25}}
\put(12.85,9.6){${\bf \times}$}
\put(13.08,7){\line(1,0){0.84}}
\put(14.92,7){\vector(-1,0){1}}
\put(15,6.9){$d_R$}
\put(12.2,6.5){$D_{17-}$}
\put(13.2,9.6){$ \left \langle {2_{17}} \right \rangle $}

\put(1,3.5){\framebox(2.3,1){$(c)$: $(M_d)_{12}$}}
\put(7.08,2){\vector(1,0){1}}
\put(8.08,2){\line(1,0){0.84}}
\multiput(9,2)(0,0.3){9}{\line(0,1){0.25}}
\put(8.85,4.6){${\bf \times}$}
\put(9.08,2){\line(1,0){0.84}}
\put(10.92,2){\vector(-1,0){1}}
\multiput(11,2)(0,0.3){9}{\line(0,1){0.25}}
\put(10.85,4.6){${\bf \times}$}
\put(11.08,2){\line(1,0){0.84}}
\put(12.92,2){\vector(-1,0){1}}
\put(2.5,1.9){$d_L$}
\put(9.2,4.6){$ \left \langle {H_d} \right \rangle $}
\put(10.2,1.5){$D_{4+}$}
\put(11.2,4.6){$ \left \langle {2_5} \right \rangle $}
\put(13,1.9){$s_R$}
\put(5.08,2){\vector(1,0){1}}
\put(6.08,2){\line(1,0){0.84}}
\multiput(7,2)(0,0.3){9}{\line(0,1){0.25}}
\put(6.85,4.6){${\bf \times}$}
\put(3.08,2){\vector(1,0){1}}
\put(4.08,2){\line(1,0){0.84}}
\multiput(5,2)(0,0.3){9}{\line(0,1){0.25}}
\put(4.85,4.6){${\bf \times}$}
\put(7.5,1.5){$Q_{7+}$}
\put(5.5,1.5){$Q_{13-}$}
\put(7.2,4.6){$ \left \langle {2_{20}} \right \rangle $}
\put(5.2,4.6){$ \left \langle {2_{13}} \right \rangle $}

\end{picture}

\caption{Froggatt-Nielsen tree graphs for $M_d$. }

\label{Fig. 2}

\end{figure}

\end{flushleft}

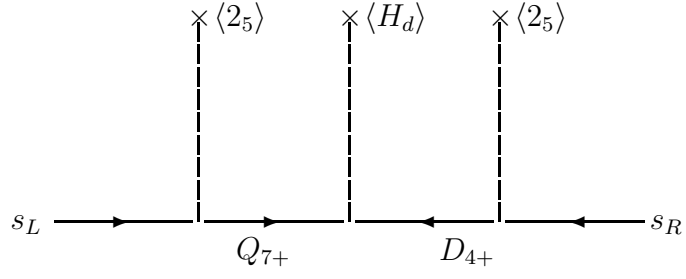
\end{document}